\documentclass[12pt]{iopart}
\usepackage{times}
\usepackage{graphicx}

\begin{document}
\title{Interpreting data from WIMP direct detection experiments}
\author{Anne M Green}
\address{Physics Department, Stockholm University, Stockholm, 106 91, SWEDEN}
\begin{abstract}
Weakly Interacting Massive Particle (WIMP) direct detection
experiments are closing in on the region of parameter space
where relic neutralinos may constitute the galactic halo dark matter.
We discuss two issues in the interpretation of data, in particular the
calculation of exclusion limits, from these experiments. Firstly we
show that the technique that has been used for calculating exclusion
limits from binned data without background subtraction produces
erroneously tight limits, and discuss alternative methods which avoid
this problem. We then argue that the standard maxwellian halo model is
likely to be a poor approximation to the dark matter distribution and
examine how halo models with triaxiality, velocity anisotropy and small
scale clumping affect exclusion limits.
\end{abstract}
\section{Introduction}

Arguably the best motivated non-baryonic dark matter candidate is the
neutralino (the lightest supersymmetric particle), and current direct
detection experiments are just reaching the sensitivity required to
probe the relevant region of parameter space~\cite{lars}.  The most
stringent exclusion limits on Weakly Interacting Massive Particles
(WIMPs) in general currently come from the Edelweiss~\cite{edelnew}
and Cryogenic Dark Matter Search (CDMS) experiments~\cite{CDMS1,CDMS2}, with
competitive constraints also having been produced by experiments which
have been optimized for double-beta decay such as
Heidelberg-Moscow (HM)~\cite{HM} and IGEX~\cite{IGEX}. The exclusion limits
from these experiments, calculated assuming a standard maxwellian
halo, are plotted in Fig. 1 along with the region of parameter space
corresponding to the DAMA collaboration's annual modulation
signal~\cite{DAMA}. Given the experimental progress, and the
tension between the DAMA collaboration's annual modulation signal and
the exclusion limits from other experiments, it is crucial to examine
the assumptions involved in interpreting data from these experiments.

We focus on two separate issues: the calculation of exclusion limits
with the correct coverage (i.e. which correspond to the stated degree
of confidence) and the effect of halo modeling on exclusion limits.
In Sec.~2 we show that the method previously used to calculate
confidence limits from experiments without background subtraction and
binned data (such as HM and IGEX) produces erroneously tight
exclusion limits, and discuss alternative criteria for calculating
exclusion limits.  In Sec.~3 we turn our attention to the
dependence of the theoretical differential event rate on the WIMP
speed distribution. We discuss the properties of galactic halos and
examine how models which reproduce these properties affect the
exclusion limits from the HM and IGEX experiments.

\begin{figure}
\begin{center}
\includegraphics{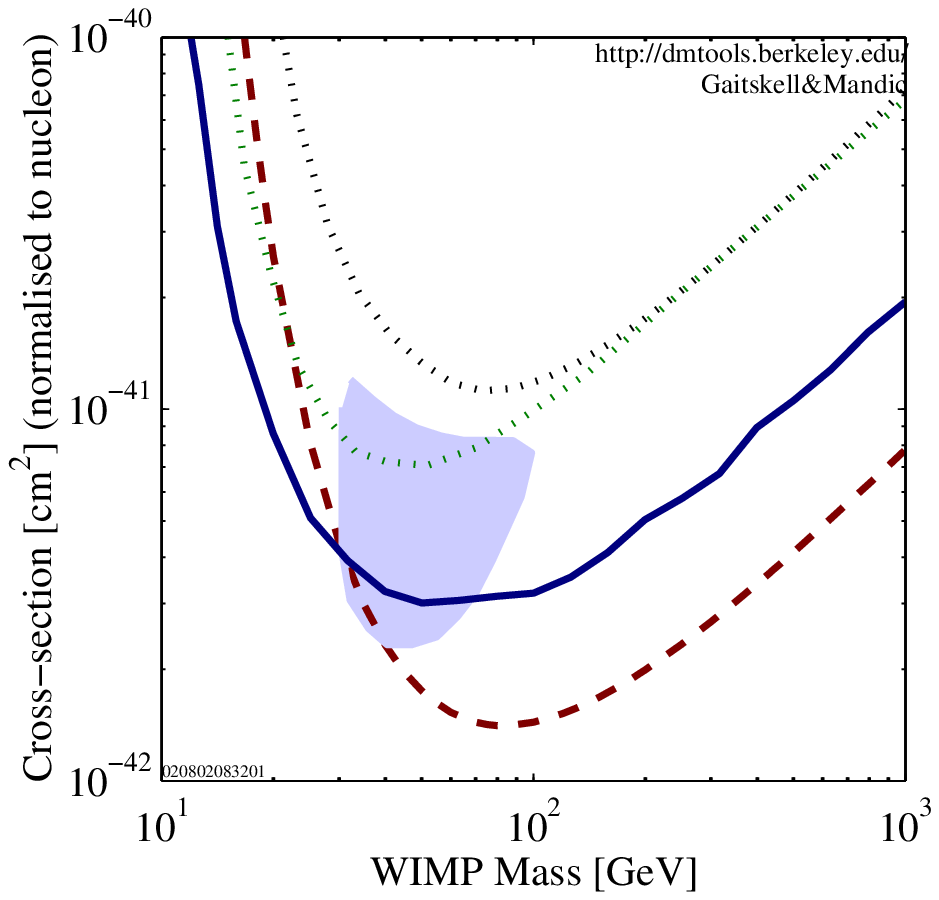}
\includegraphics{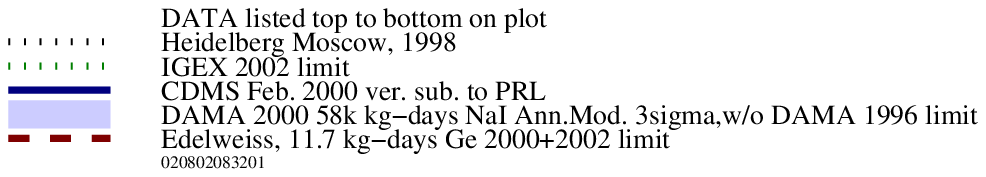}
\end{center}
\caption{Exclusion limits, and the DAMA annual modulation region, from
the WIMP direct detection experiments discussed in the text, assuming
a standard maxwellian halo model, plotted using the interactive limit
plotter at http://dmtools.berkeley.edu/limitplots/ .}
\end{figure}

\section{Calculating exclusion limits}
In experiments without background subtraction, such as
Heidelberg-Moscow (HM)~\cite{HM} and IGEX~\cite{IGEX}, any set of WIMP
parameters (mass and cross-section) which would produce more events
than are observed, at some confidence level, can be excluded at that
confidence level. The HM and IGEX collaborations have data in bins of
width 1 keV (42 bins from 9 keV for HM and 45 bins from 4 keV for
IGEX) and in their analysis~\cite{HM,IGEX} the 90$\%$ confidence
exclusion limit on the cross-section, $\sigma_{{\rm p}}$, is
calculated by finding for each $m_{\chi}$ the value of $\sigma_{{\rm
p}}$ for which the theoretical event rate in {\em one} of the bins
just exceeds the 90$\%$ confidence limit on the event rate in that
bin. By definition a 90$\%$ upper confidence limit means that there is
a 10$\%$ probability that a theoretical number of events greater than
the 90$\%$ confidence limit produced the observed number of events,
{\em for any given bin}.

We will now consider an ensemble of bins,
where $N_{{\rm t}}$ is the total number of bins, $N_{{\rm e}}$ is the
number of bins for which the theoretical number of events exceeds the
90$\%$ bin confidence limit and $P(N_{{\rm e}})$ is the probability
distribution of $N_{{\rm e}}$. For an ensemble of bins
\begin{eqnarray}
\label{cl}
P(N_{{\rm e}} > 0) & =&  1 - P(N_{{\rm e}}=0) = 1- (0.9)^{N_{{\rm t}}} 
              \nonumber \\
                     & >& 0.1  \,\,\,\,\,  {\rm if} \, N_{{\rm t}} > 1 \,.
\end{eqnarray}
In other words the exclusion limits found using the standard analysis
actually correspond to a lower degree of confidence than 90$\%$, and
are hence erroneously tight. 

The probability distribution of the number of bins $N_{{\rm e}}$
exceeding their $100 p \%$ (where $0 < p < 1$) bin confidence limit
can be used to formulate criteria which produce true 90$\%$ minimum
confidence exclusion limits (see Ref.~\cite{statpap} for further
details).  Firstly, for any given total number of bins, we can vary
$p$ and find the confidence level $cl$ for which the probability that
none of the bins exceed their $100 cl \%$ bin confidence is
90$\%$. For $N_{{\rm t}}= 45$, $ cl = 0.9977$. We can also take
$p=0.9$ and calculate the 90$\%$ minimum upper confidence limit on the
number of bins which exceed their 90$\%$ bin confidence limit.  For
$N_{{\rm t}}=45$, $P(N_{{\rm e}} < 8) = 0.924 $.  Since $N_{{\rm e}}$
can only take on integer values this criteria does not produce exact
90$\%$ exclusion limits, but the amount by which the confidence limits
are stronger than 90$\%$ is known, and fixed for fixed $N_{{\rm t}}$.
A simple minded way to avoid the problem of calculating correct
overall confidence limits would be to discard all but one of the
energy bins; $N_{{\rm t}}=1$ then and the 90$\%$ bin confidence limit
gives an overall 90$\%$ confidence limit. The obvious choice for the
energy bin to use is the lowest, threshold, energy bin, since ${\rm d}
R/ {\rm d} E_{{\rm R}}$ decreases roughly exponentially with
increasing $E_{{\rm R}}$ for the standard Maxwellian halo model so
that this bin is often the most constraining. This wasteful method
will of course produce weaker exclusion limits than can be found using
the entire data set though.

The exclusion limits resulting from these criteria, assuming a
standard maxwellian halo, are plotted in Fig.~2 for the IGEX 60 kg-day
data~\cite{IGEXold}. We see that the standard technique produces
exclusion limits that are erroneously tight (by roughly $\Delta (
\log_{10} \sigma_{{\rm p}} ) \sim 0.2.$ for $m_{\chi} > 80$ GeV).  No
more than 7 bins exceeding their 90$\%$ bin confidence limit produces
exclusion limits which are especially weak for small $m_{\chi}$ and
does not produce exactly 90$\%$ exclusion limits, since the number of
bins is an integer. Only using one of the energy bins is wasteful and
produces substantially weaker exclusion limits than could be found
using the entire data set. Requiring that no bin exceeds its 99.75$\%$
bin confidence limit therefore appears to be the best of these methods
for producing genuine 90$\%$ overall confidence limits.

Several other techniques have been proposed for dealing with this
problem. The CRESST collaboration~\cite{CRESST} carry out a maximum
likelihood fit of their measured signal with an empirical function to
effectively find the maximum WIMP signal which can be `hidden' behind
the background. Yellin~\cite{yellin} has devised an `optimal interval
method' for use with low background unbinned data, which has been used
by the CDMS collaboration to calculate a limit from their data without
Monte Carlo subtraction of the neutron background~\cite{CDMS2}. This
method effectively chooses the binning of the data which gives the
tightest exclusion limit, while taking into account the statistical
`penalty' associated with the freedom in the binning.

\begin{figure}
\begin{center}
\includegraphics[width=0.6\textwidth]{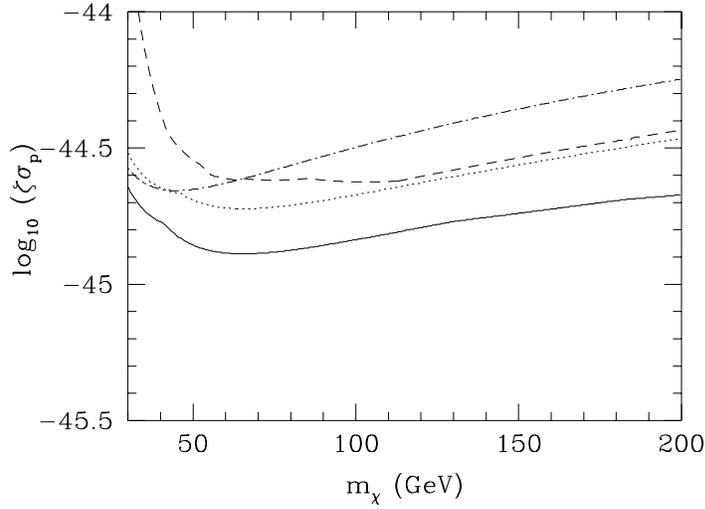}
\end{center}
\caption{The spin-independent WIMP cross-section exclusion limits
found from the IGEX 60 kg-day data~\cite{IGEXold}, assuming the
standard maxwellian halo model, from: using the standard no bin
exceeding its 90$\%$ bin confidence limit method (solid), no bin
exceeding its 99.77$\%$ bin confidence limit (dotted), no more than 7
bins exceeding their 90$\%$ bin confidence limit (dashed) and
the lowest energy bin not exceeding its 90$\%$ bin confidence limit
(dot-dashed).}
\end{figure}

\section{Milky Way halo modeling}
The direct detection event rate and its energy distribution are
determined in part by the WIMP speed distribution (see eq.~3
below). Data analyzes nearly always assume a standard smooth halo
model with isotropic maxwellian velocity distribution.  Belli
et. al.~\cite{damare} have recently reanalyzed the DAMA collaboration's
annual modulation signal~\cite{DAMA} for a range of halo models,
finding that the allowed region of WIMP mass--cross-section parameter
space is significantly enlarged.  Models with triaxiality or velocity
anisotropy may produce a significant change even in the mean
differential event rate~\cite{newevans,uk}.  Furthermore all of the
non-standard halo models which have previously been considered are
essentially smooth, while N-body simulations produce dark matter
halos which contain significant amounts of substructure~\cite{Nbody1}.

We first review the global and local properties of dark matter halos
and then examine how these properties may affect the calculation of
exclusion limits.

\subsection{Review of halo properties}

\subsubsection{Global properties}

Observational constraints on the structure of dark matter halos depend
on the relation of luminous tracer populations to the underlying
density distribution, and are complicated by galactic structure and
projection effects (see e.g. Ref.~\cite{sackett}).  Given the
difficulties involved in `observing' dark matter halos it makes sense
to turn to numerical simulations for information on their possible
properties.  In CDM cosmologies structure forms hierarchically, from
the top down~\cite{cdmgen}. Small objects (often known as subhalos)
form first, with larger objects being formed progressively via mergers
and accretion. The internal structure of large galaxy size halos is
determined by the dynamical processes which act on the accreted
components, for instance the tidal field of the main halo can strip
material away from a subhalo~\cite{bt,hay} producing tidal streams
along its orbit~\cite{ts}.

Current simulations of galaxy halos within a
cosmological context can resolve sub-kpc scales (see
e.g.~\cite{Nbody1,Nbody2}). Discrepancies between the halos produced
in these simulations (which have lots of surviving dwarf galaxy sized
subhalos and steep central profiles) and observations, have led to
claims of a crisis for the cold dark matter model (see
Ref.~\cite{cdmcrisis} and references therein for an extensive
discussion). Most relevant for the local dark matter distribution is
the subhalo problem which may be, at least partly, due to
complications in comparing the observed luminous matter with the dark
matter distribution from the simulations. In particular it has been
argued that gas accretion onto low mass halos may be inhibited after
reionization so that a large fraction of the subhalos remain
dark~\cite{sf}. It has also been shown that if the observed dwarf
galaxies themselves have dark halos, then their masses have been
underestimated and correcting for this would go toward resolving the
discrepancy with observations~\cite{hay,dg}. The survival of subhalos
is at least partly due to their concentrated profiles, so any
modification to the simulations which produced halos with shallower
central profiles could also reduce the number of surviving
subhalos, however. Despite the ongoing debate regarding the detailed comparison
of the small scale properties of simulated halos with observations,
cosmological simulations may still provide us with useful information
about the global properties of galactic halos.

The shape of simulated halos varies, not just between different halos
of the same mass, but also as a function of radius within a single
halo, strongly if the halo has undergone a major merger relatively
recently. Two high resolution Local Group halos studied in detail in
Ref.~\cite{mooredm} have axis ratios of $1:0.78:0.48$ and
$1:0.45.0.38$ at the solar radius and $1:0.64:0.40$ and $1:0.87:0.67$
at the virial radius. Adding dissipative gas to simulations tends to
preserve the short-to-long axis ratio while increasing the
intermediate-to-long axis ratio~\cite{gas}.

The anisotropy parameter $\beta(r)$, defined as
\begin{equation}
\label{defbeta}
\beta(r)= 1 - \frac{<v_{ \theta}^2>+<v_{ \phi}^2> }{2 <v_{{\rm r}}^2>} \,,
\end{equation}
where $<v_{{\rm \theta}}^2>$, $<v_{{\rm \phi}}^2>$ and $<v_{{\rm
r}}^2>$ are the means of the squares of the velocity components, also
varies with radius.  Typically $\beta(r)$ grows, although not
monotonically, from roughly zero in the center of the halo to close to
one at the virial radius, with non-negligible variation between halos
(see Fig. 23 of Ref.~\cite{fm}). The high resolution galactic mass
halos studied in Ref.~\cite{moore99} have $\beta(R_{\odot})$ in the
the range 0.1-0.4, corresponding to radially biased orbits.

This broad picture, that galaxy halos are triaxial with anisotropic
velocity distributions, is supported by various observations (see
e.g. Refs.~\cite{sackett,om2,dsg,glob}).

\subsubsection{Local dark matter distribution}

The local dark matter distribution, which is crucial for direct
detection experiments, can not be probed directly by cosmological
simulations; the smallest subhalos resolvable in the highest
resolution simulations have mass of order $10^{7} M_{\odot}$ and it is
not possible to fully resolve substructure within subhalos. Little
substructure is found within the central regions of simulated halos,
however it is not known whether the subhalos have been destroyed by
tidal stripping or if this is purely a resolution
effect~\cite{mooredm}. This is relevant for the local dark matter
distribution as the solar radius ($R_{\odot} \approx 8$ kpc) is small
compared with the radius of the MW halo. The computing power required
to directly probe the local dark matter distribution will probably not
be available for a decade or so~\cite{mooredm}, therefore other
numerical~\cite{mooredm,hws} and semi-analytic~\cite{swf} approaches
have been used to address the problem. This work is reviewed in
detail in Ref.~\cite{greennew}. In summary, there is currently no
consensus on the local dark matter velocity distribution, with the
results obtained depending on the method used to extrapolate to small
scales below the resolution limit of the cosmological simulations.  The
local dark matter velocity distribution may be well approximated by a
smooth multi-variate gaussian, with clumps of high velocity particles
present if the MW halo has undergone a recent major merger~\cite{hws},
on the other hand it is possible that the local dark matter density
could be zero or that a single dark matter stream with small velocity
dispersion could dominate or that many tidal streams could overlap to
give a smooth distribution~\cite{mooredm}.

\subsection{Effect on exclusion limits}
The differential WIMP event rate due to scalar interactions can be
written in terms of the WIMP scattering cross section on the proton,
$\sigma_{{\rm p}}$~\cite{jkg}:
\begin{equation}
\frac{{\rm d} R}{{\rm d}E} = \zeta \sigma_{{\rm p}} 
              \left[ \frac{\rho_{0.3}}{\sqrt{\pi} v_{0}}
             \frac{ (m_{{\rm p}}+ m_{\chi})^2}{m_{{\rm p}}^2 m_{\chi}^3}
             A^2 T(E) F^2(q) \right] \,,
\end{equation}
where the local WIMP density, $\rho_{\chi}$ is normalized to a
fiducial value $\rho_{0.3} =0.3 \, {\rm GeV/ cm^{3}}$, such that
$\zeta=\rho_{\chi} / \rho_{0.3}$, $m_{A}$ is the atomic mass of the
target nuclei, $E$ is the recoil energy of the detector nucleus, and
$T(E)$ is defined as~\cite{jkg}
\begin{equation}
\label{tq}
T(E)=\frac{\sqrt{\pi} v_{0}}{2} \int^{\infty}_{v_{{\rm min}}} 
            \frac{f_{v}}{v} {\rm d}v \,,
\end{equation}
where $f_{v}$ is the WIMP speed distribution in the rest frame of the
detector, normalized to unity, and $v_{{\rm min}}$ is the minimum
detectable WIMP speed
\begin{equation}
v_{{\rm min}}=\left( \frac{ E (m_{\chi}+m_{A})^2}{2 m_{\chi}^2 m_{A}} 
             \right)^{1/2} \,.
\end{equation}

We will first examine the change in the WIMP speed distribution, and
hence the exclusion limits, for triaxial and anisotropic, but still
smooth, halo models. To date two self-consistent triaxial and/or
anisotropic halo models have been studied in relation to WIMP direct
detection: the logarithmic ellipsoidal model~\cite{newevans} and the
Osipkov-Merritt anisotropy model~\cite{OM}, studied in
Ref.~\cite{uk}. We extend the previous work by focusing on parameters
which span the range of observed and simulated halo properties.

\begin{figure}
\label{figf}
\begin{center}
\includegraphics[width=0.6\textwidth]{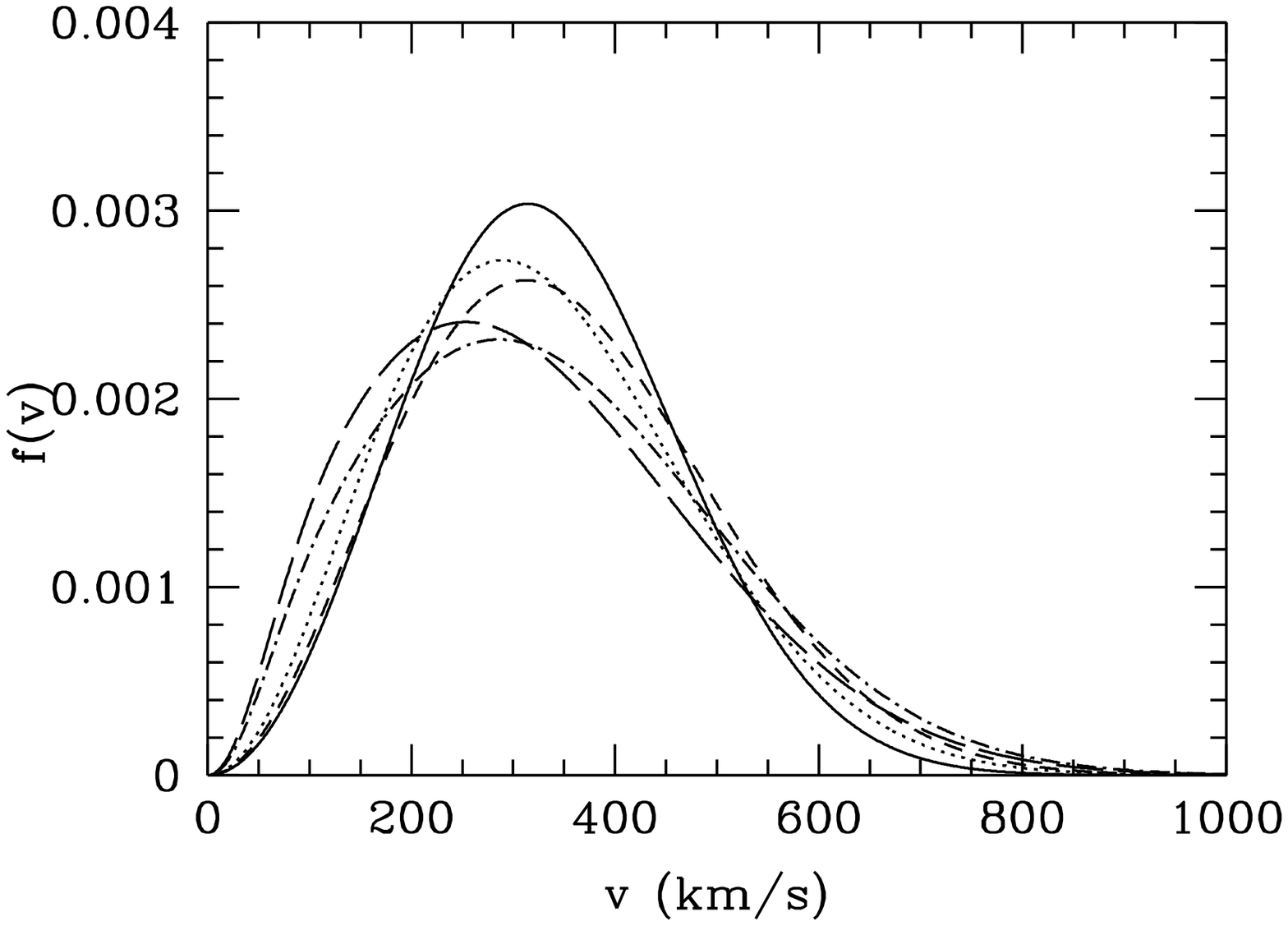}
\includegraphics[width=0.6\textwidth]{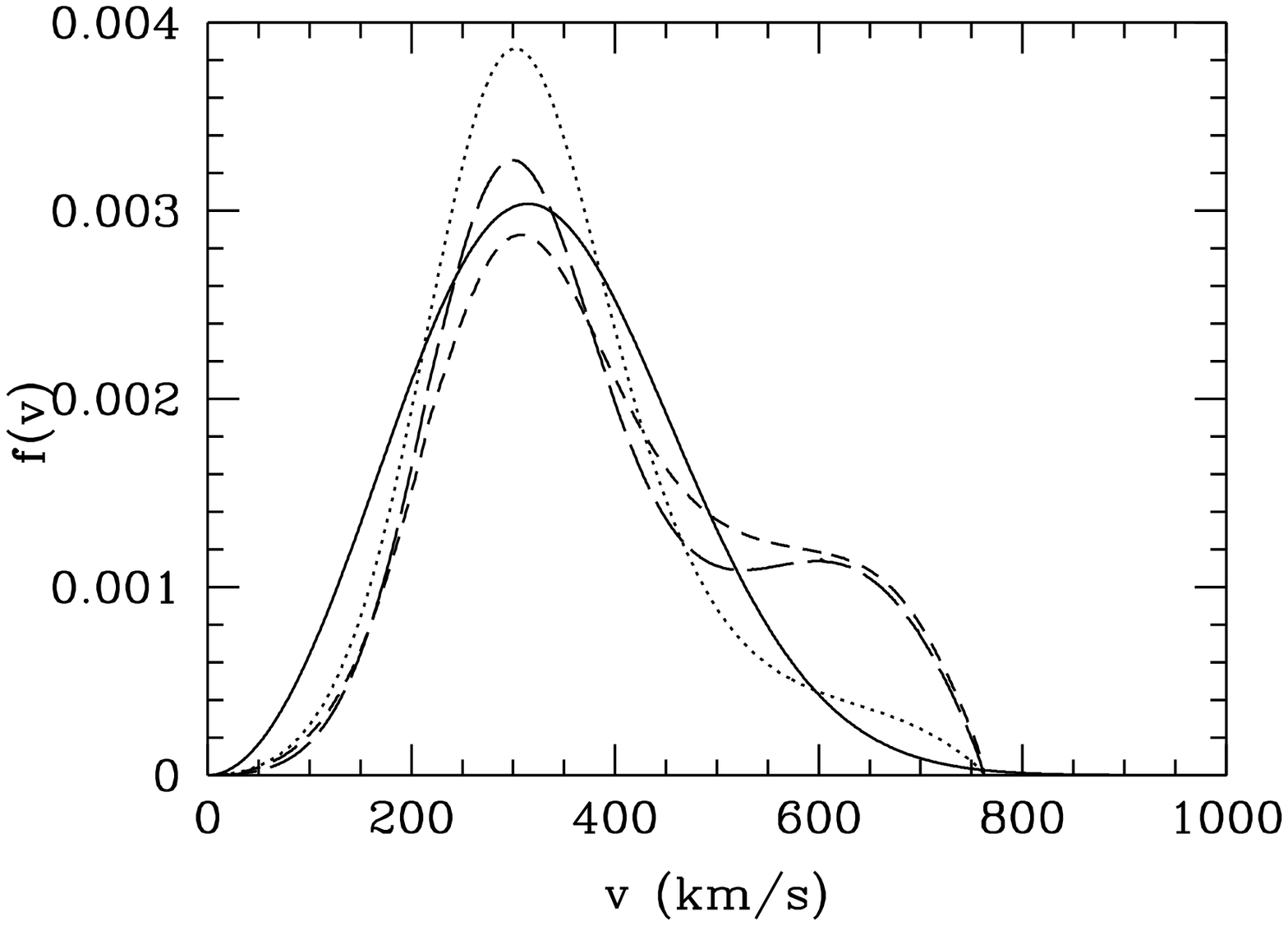}
\end{center}
\caption{The speed distributions, in the rest frame of the Sun, for
the standard halo model (solid line), and the logarithmic ellipsoidal
model on the intermediate axis (upper panel) for p=0.9, q=0.8 and
$\beta=0.1/0.4$ (dotted/short dashed) and for p=0.72, q=0.7 and
$\beta=0.1/0.4$ (long dashed/dot dashed) and for the OM anisotropy
model (lower panel) with $\beta=0.13, 0.31$ and $0.4$ (dotted,
short-dashed, and long-dashed). }
\end{figure}

The logarithmic ellipsoidal model~\cite{newevans} is the simplest
triaxial generalization of the isothermal sphere and the velocity
distribution can be approximated by a multi-variate gaussian on either
the long or the intermediate axis \footnote{Of course there is no
reason to expect the Sun to be located on one of the axes of the
halo.}.  We consider parameter values $p= 0.9, q=0.8$ corresponding to
axial ratios $1:0.78:0.48$ and $p=0.72, q=0.70$ corresponding to
$1:0.45:0.38$, and values of the anisotropy parameter which give
$\beta=0.1$ and $0.4$. The speed distributions on the intermediate
axis, in the rest frame of the Sun normalized to unity, are plotted
in Fig. 3 along with that for the standard maxwellian halo model. For
both positions the triaxial models have a wider spread in speeds than
the standard model, so that the differential event rate will decrease
less rapidly with increasing recoil energy, but the change is small on
the major axis. This is because the change in the speed distribution
is largely determined by the velocity dispersion in the $\phi$
direction.  On the major axis, for parameter values which give $0.1 <
\beta < 0.4 $, all three components of the velocity have roughly the
same dispersion, whereas on the intermediate axis the velocity
dispersion in the $\phi$ direction is significantly larger than that
in the $z$ direction.

\begin{figure}
\begin{center}
\label{IGEX}
\includegraphics[width=0.6\textwidth]{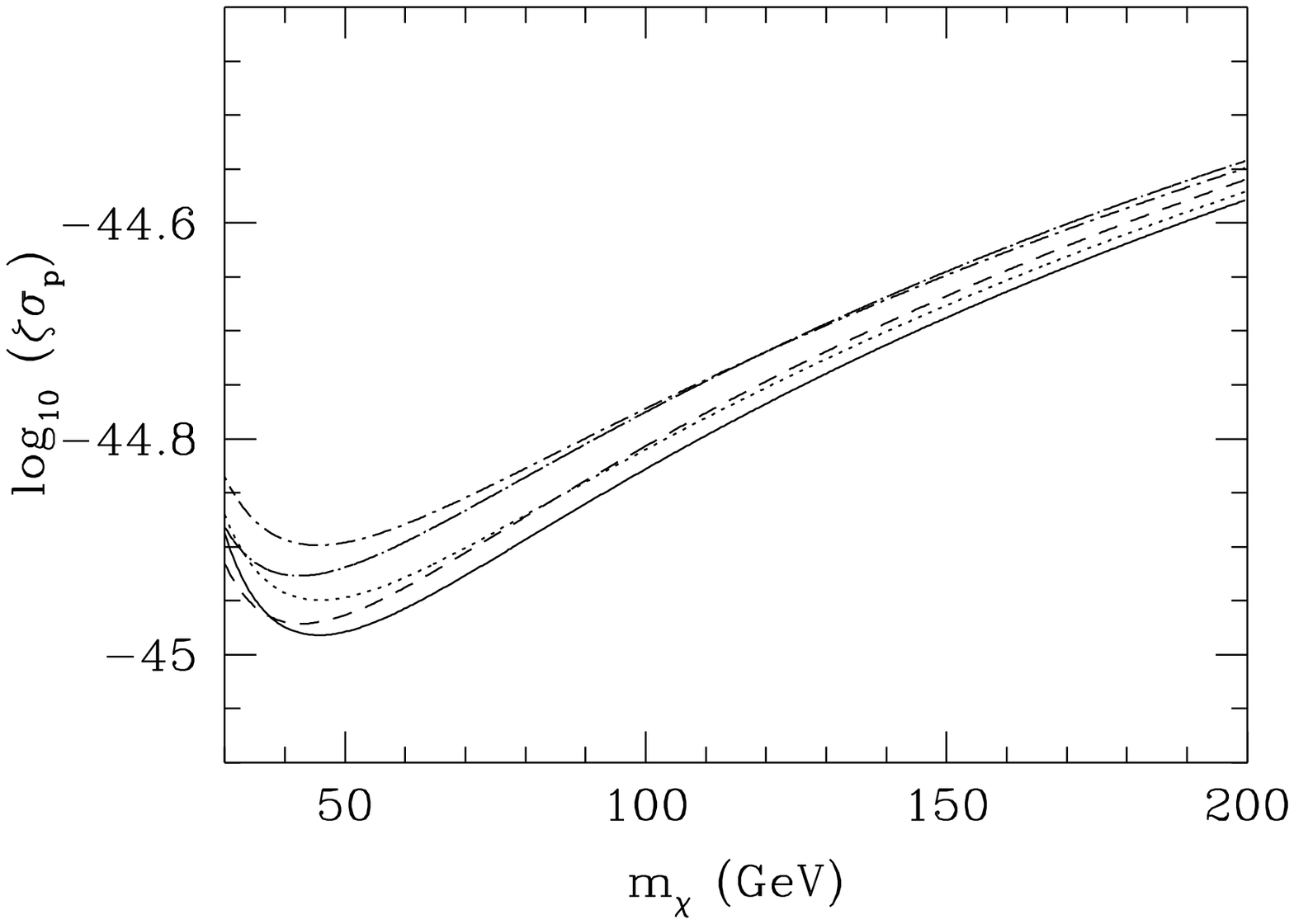}
\includegraphics[width=0.6\textwidth]{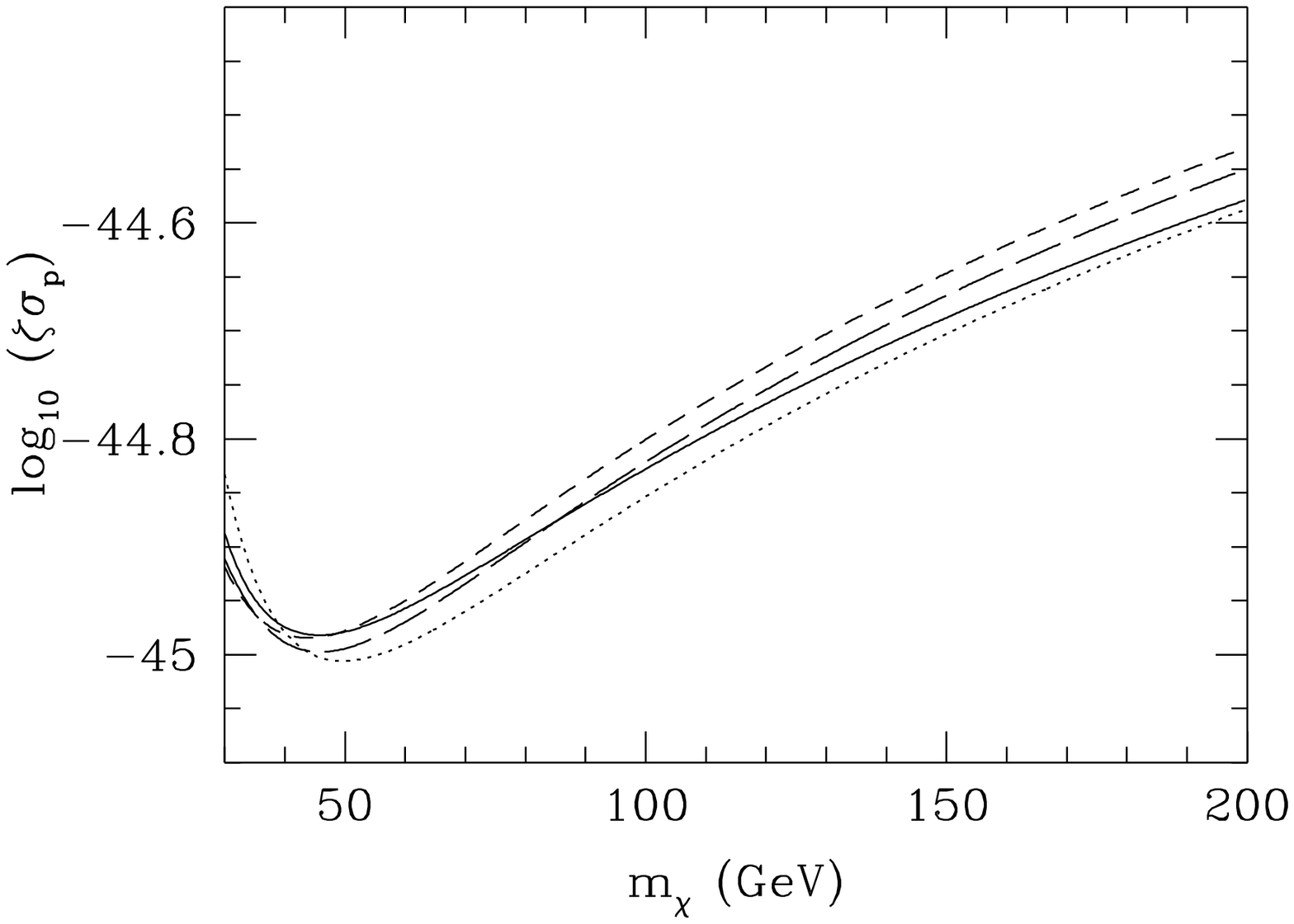}
\end{center}
\caption{The exclusion limits from the IGEX
experiment for the logarithmic ellipsoidal model, location on the
intermediate axis (upper panel) and for the OM model
(lower panel). Parameters and line types as in Fig. 3.}
\end{figure}

In the Osipkov-Merritt (OM) model, which assumes a spherically
symmetric density profile, the velocity anisotropy varies as a
function of radius as
\begin{equation}
\beta(r) = \frac{r^2}{r^2 + r_{{\rm a}}^{2}} \,,
\end{equation}
so that the degree of anisotropy increases with increasing radius, as
is found in numerical simulations. Following Ref.~\cite{uk} we assume
a NFW~\cite{NFW} density profile with scale radius $r_{{\rm s}}=20$ kpc. We
use values of the anisotropy radius $r_{{\rm a}}=20, 12,$ and $9.8$
kpc which correspond to $\beta(R_{0}) = 0.14, 0.31$ and $0.4$
respectively. The resulting speed distributions are plotted in
Fig. 3. The excess at large $v$ is due to the increased number of
particles on very elongated, nearly radial orbits~\cite{uk}.

To assess the effect of changes in the speed distribution on exclusion
limits we need to take into account the detector response, including
the difference between the observed energy of an event and the actual
recoil energy, non-zero energy threshold and energy resolution (see
Ref.~\cite{ls} for further details), as these factors may blur out the
effects of changes in the speed distribution. In Fig. 4 we plot the
exclusion limits found from the IGEX data by requiring that the data
in no more than one energy bin exceeds its 99.77$\%$ confidence limit,
so as to produce 90$\%$ overall confidence limits (see above and
Ref.~\cite{statpap}), for the logarithmic ellipsoidal model and for
the OM anisotropy model. We also plot the exclusion limits from the
Heidelberg-Moscow (HM) experiment~\cite{HM} for the OM anisotropy
model in Fig. 5. Comparing Figs. 4 and 5 we see that the change in the
exclusion limits depends not only on the halo model under
consideration, but also on the data being used; for IGEX the change in
the exclusion limits is largest for large $m_{\chi}$, while for HM the
change is largest for small $m_{\chi}$. For different $m_{\chi}$,
different energy ranges can be most constraining; for the IGEX data
the lowest energy bin is always the most constraining, while for HM as
 $m_{\chi}$ increases the constraint comes from higher energy
bins. It should therefore be borne in mind when comparing exclusion
limits from different experiments, that changing the assumed WIMP
speed distribution will affect the limits from different experiments
differently.

\begin{figure}
\begin{center}
\label{HM}
\includegraphics[width=0.6\textwidth]{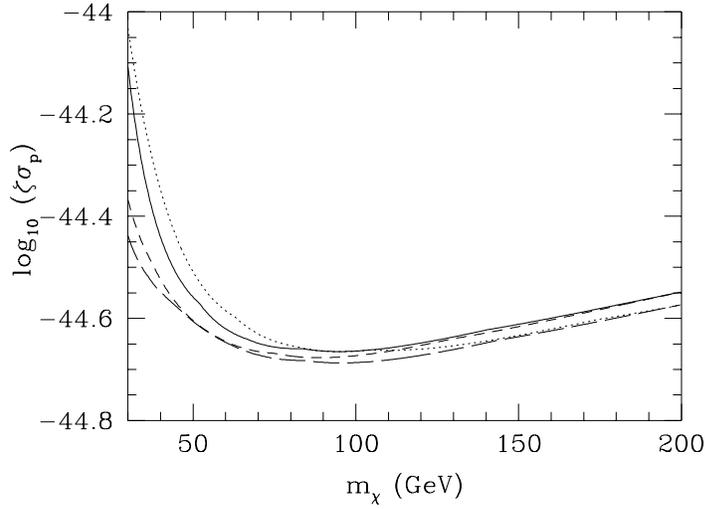}
\end{center}
\caption{The exclusion limits from the HM experiment for the OM
model. Parameters and line types as in the lower panel of Fig. 3.}
\end{figure}

The change in the exclusion limits is not huge (of order tens of
per-cent) for the experiments we have considered, however these
experiments are not optimized for WIMP detection. The change in the
differential event rate, and hence the exclusion limit, would be
significantly larger for an experiment with better energy resolution
and lower threshold energy (see Ref.~\cite{greennew}). We have also
seen that different models with the same value for the anisotropy
parameter $\beta$ have very different speed distributions, and hence a
different effect on the exclusion limits . Furthermore it is
conceivable that the local WIMP speed distribution may deviate even
further from the standard maxwellian distribution than the models
that we have considered.

\vspace{1.0cm}

We will now turn our attention to the possibility that the local dark
matter distribution is not smooth.  Even if dynamical processes
produce a smooth background dark matter distribution, late accreting
clumps may lead to a local density enhancement and velocity
clumping~\cite{swf,hws}, producing a shoulder in the differential
event rate, if their density and velocity with respect to the earth
are large enough~\cite{swf}. For the experiments we have been
considering the lower energy bins are most constraining, so that only
very rare high density and velocity streams would have a
non-negligible effect on the exclusion limits. The effect of these
late accreting clumps on the annual modulation and directional signals
would be more significant however~\cite{swf}.

Finally we look at the consequences of the more speculative
possibility that small subhalos may survive at the solar radius. We
could then be located within a subhalo with local density in excess of
the mean value of $0.3 \, {\rm GeV cm^{-3}}$, on the other hand it is
even possible that we could be in a region between clumps and streams
where the WIMP density is zero~\cite{mooredm}. In the latter case all
attempts at WIMP direct detection would be doomed to failure, and
exclusion limits would tell us nothing about the WIMP
cross-section. At the other extreme a tiny subhalo at the earth's
location would produce a distinctive signal and, due to its high
density, an enhanced event rate, making it easier to detect WIMPs of a
given cross-section. Subhalos with $M \ll 10^{9} M_{\odot}$ would have
negligible velocity dispersion and hence a delta-function speed
distribution~\cite{swf}. The resulting theoretical differential event
rate would be a step function with amplitude inversely proportional to
the speed of the subhalo with respect to the earth, and position
increasing with increasing relative speed and WIMP mass. Consequently
for small subhalo velocities and WIMP masses there would be no
constraint on the WIMP cross-section (no WIMPs would have sufficient
energy to cause an observable recoil), but as the WIMP mass is
increased the constraints would become much tighter as then all the
WIMPs would be energetic enough to cause events of a given recoil
energy.

\vspace{1.0cm}

In summary, triaxiality and velocity anisotropy lead to non-negligible
changes in the exclusion limits, even for non-optimized
detectors. Furthermore the changes are different for different data
sets and depend on how the anisotropy is modeled. If the local WIMP
distribution is dominated by small scale clumps then the density in
the solar                              neighborhood may be zero (making it impossible to detect
WIMPs) or significantly enhanced (making it easier to detect WIMPs
with a given cross-section), and the exclusion limits are changed
dramatically. Clearly the survival of subhalos at the solar radius is
a very important issue for WIMP direct detection.

\section{Conclusions}

We have examined two aspects of the interpretation of data from WIMP
direct detection experiments. We have seen that care needs to be taken
when calculating exclusion limits from experiments without background
subtraction so as to produce correct limits. Yellin's optimal interval
method provides a sophisticated and robust solution to this problem
for experiments with unbinned data and relatively small numbers of
events~\cite{yellin}, in other cases Poisson statistics can be used to
formulate criteria which produce correct limits~\cite{statpap}.

We have also seen that even if the local WIMP distribution is smooth
its speed distribution may deviate significantly from the standard
maxwellian, and this has a non-negligible effect on exclusion limits
and, crucially, effects the limits from different experiments
differently. Constraints (and in the future possibly best fits)
calculated assuming a standard maxwellian halo could be
erroneous~\cite{uk}. The derivation of reliable constraints on WIMP
parameters and comparison of results for different experiments
requires a theoretical framework for dealing with the uncertainty in
the WIMP speed distribution. On the other hand, more optimistically,
it might be possible to derive useful information about the local
velocity distribution, and hence the formation of the galactic halo,
if WIMPs were detected~\cite{swf,hws}.

\end{document}